\documentclass[english,3p]{elsarticle}
\usepackage[T1]{fontenc}
\usepackage[latin9]{inputenc}
\usepackage{amsmath}
\usepackage{amssymb}
\usepackage{graphicx}

\makeatletter
\newtheorem{theorem}{Theorem}

\newtheorem{proposition}{Proposition}

\newtheorem{definition}{Definition}
\newtheorem{fact}{Fact}

\makeatother

\usepackage{babel}
\begin{document}

\title{\textbf{Tensor product extension of entanglement witnesses and their
connection with measurement device independent entanglement witness.}}

\author[IF,KCIK]{A. Rutkowski\corref{cor1}}
\ead{arut@fizyka.umk.pl}
\cortext[cor1]{Corresponding author}
\author[PG,KCIK]{P. Horodecki}

\address[IF]{Institute of Theoretical Physics and Astrophysics,
University of Gda\'{n}sk, 80-952 Gda\'{n}sk, Poland}
\address[KCIK]{National Quantum Information Center, 81-824 Sopot,
Poland}
\address[PG]{Faculty of Applied Physics and Mathematics, Gda\'{n}sk
University of Technology, 80-233 Gda\'{n}sk, Poland\\}

\begin{abstract}
We provide a new extension of entanglement witnesses for $\mathbb{C}^{d_{1}}\otimes\mathbb{C}^{d_{2}}$
systems. Our construction preserves the properties of indecomposability
and spanning property of entanglement witnesses. We show how our concept
of extended entanglement witnesses is connected with the idea of measurement device
independent entanglement witnesses.
\end{abstract}
\maketitle
%\pacs{03.67.Mn,03.65.Ud}

%\maketitle

\section{Introduction}

Detection of entanglement is one of the fundamental problem in quantum information theory \citep{QIT} . It is well known
that it is extraordinary hard to check whether for a given density matrix
describing a quantum state of the complex system is separable or
entangled. There exist  several operational criteria which allow us 
to detect quantum entanglement (see e.g. \citep{Horodecki-review}). One of  the most famous criterion is based on the
partial transposition: if a state $\rho$ is separable then its partial
transposition $\rho^{\Gamma}=(\mathbb{I}\otimes{\rm T})\rho$ is positive
\citep{HHH}. States which are positive under partial transposition
are called PPT states. It is easy to see that each separable state is necessarily
PPT but the converse is not always true. The most general approach to characterize
quantum entanglement uses the notion of an entanglement witness (EW)
\citep{HHH,Terhal1,Terhal2}. It turns out that a state is entangled
if and only if it is detected by some EW \citep{HHH}. There was a
considerable effort in constructing and analyzing the structure of
EWs \citep{O}--\citep{iran}. However, there is no general method
to construct such objects. An entanglement witness which detects a maximal set of entanglement
is said to be optimal, as was introduced in Ref. \citep{O}. Unfortunately,
there is no complete characterization of  indecomposable optimal EW
and only very few examples of optimal indecomposable optimal EW
are available in the literature. Optimal EWs are of primary importance
since to perform complete classification of quantum states of a bipartite
system it is enough to use only optimal EWs. In the present paper
we provide some extension of EWs for $\mathbb{C}^{d_{1}}\otimes\mathbb{C}^{d_{2}}$
systems which preserve crucial properties, namely indecomposability
and optimality. We also show how  our concept is connected with the
idea of measurement  device independent entanglement witnesses (MDIEW) \citep{Gisin,Buscemi}.
The latter are the witnesses that are robust against misalignment of the 
measurement setup or even the degrees of freedom of the system entanglement of which is to be tested.
We prove that our extensions provide alternative way of explanation of the nature of MDIEW.

\section{Main results} First, we introduce  some basic terminology of entanglement witnesses and states used in this letter. Let $\mathcal{H}$ be a separable complex Hilbert space and 
 $\mathcal{B}\left(\mathcal{H}\right)$ be the algebra of all bounded linear operators on $\mathcal{H}$ . We can define the set:  

\[
\mathcal{S}\left(\mathcal{H}\right)=\left\{ \rho\in\mathcal{B}\left(\mathcal{H}\right)|\rho\geq0,\,\text{Tr}\rho=1\right\} ,
\]
i.e the set of all states on $\mathcal{H}$. If $\mathcal{H}$ and
$\mathcal{K}$ are finite dimensional, a state in the bipartite composition
system $\rho\in\mathcal{S}\left(\mathcal{H}\otimes K\right)$ is said
to be separable if can be written as $\rho=\sum_{i=1}^{k}p_{i}\rho_{i}\otimes\sigma_{i}$,
where $\rho_{i}$ and $\sigma_{i}$ are states on $\mathcal{H}$ and
$\mathcal{K}$, respectibely, and $p_{i}$ are positive numbers with
$\sum_{i=1}^{k}p_{i}=1.$ Otherwise, $\rho$ is said to be inseparable
or entangled.

 One of the most general approaches to characterize quantum entanglement uses a notion of an entanglement witness. The term of  entanglement witness was introduced first time in \citep{Terhal1}. Entanglement witness  allows us to detect  quantum states without  full information about this state.  Every entanglement witness detects something \citep{O}, since it detects in particular the projector on the subspace corresponding to the negative eigenvalues. Let us recall the definition of entanglement witness.

\begin{definition}A Hermitian operator $\mathcal{W}_{AB}\in\mathcal{B}(\mathbb{C}^{d_{A}}\otimes\mathbb{C}^{d_{B}})$
defined on a tensor product $\mathcal{H}=\mathcal{H}_{A}\otimes\mathcal{H}_{B}$
is called an EW iff

\begin{equation}
\mbox{Tr}(\mathcal{W}_{AB}\sigma_{AB})\geq0,\label{eq:sepcond}
\end{equation}
for all separable states $\sigma_{{\rm AB}}\in\mathcal{S}(\mathbb{C}^{d_{A}}\otimes\mathbb{C}^{d_{B}})$,
and here exists an entangled state $\rho_{AB}\mathcal{\in S}(\mathbb{C}^{d_{A}}\otimes\mathbb{C}^{d_{B}})$
such that

\begin{equation}
\mbox{Tr}(\mathcal{W}_{AB}\rho_{AB})<0,
\end{equation}
(one says that $\rho_{AB}$ is detected by $\mathcal{W}_{AB}$). \end{definition}
We know that for a given operator $\mathcal{W}_{AB}$ it is extremely
hard to check whether it is an  entanglement witness or not. Here, we shall give a method to cunstruct a new witness by the use of the known witness.

\begin{theorem} If $\mathcal{W}_{AB}\in\mathcal{B}(\mathbb{C}^{d_{A}}\otimes\mathbb{C}^{d_{B}})$
is entanglement witness, then so is  $\mathcal{W}_{A'ABB'}\in\mathcal{B}(\mathbb{C}^{d_{A'}}\otimes\mathbb{C}^{d_{A}}\otimes\mathbb{C}^{d_{B}}\otimes\mathbb{C}^{d_{B'}})\simeq\mathcal{B}(\mathbb{C}^{d_{A'}d_{A}}\otimes\mathbb{C}^{d_{B}d_{B'}})$  for any $d_{A'},d_{B'}=1,.\ldots,d\leq\propto \label{eq:witness}$,
where

\begin{equation}
\mathcal{W}_{A'ABB'}=\mathrm{P}_{A'}\otimes\mathcal{W}_{AB}\otimes\mathrm{P}_{B'},
\end{equation}
and $\mathrm{P}_{A'}\in\mathcal{B}(\mathbb{C}^{d_{A'}}),\mathrm{P}_{B'}\in\mathcal{B}(\mathbb{C}^{d_{B'}})$
are any positive semidefinite operators. \end{theorem}

$\mathit{Proof.}$ We need to show that $\mbox{Tr}(W_{A'ABB'}\sigma_{A'ABB'})\geq0,$
for any separable state $\sigma_{A'ABB'}\in\mathcal{S}(\mathbb{C}^{d_{A'}d_{A}}\otimes\mathbb{C}^{d_{B}d_{B'}})$. We use a spectral decomposition of operators $\mathrm{P}_{A'}={\textstyle \sum}_{i=1}^{d_{A'}}\lambda_{i}E_{i}$
and $\mathrm{P}_{B'}={\textstyle \sum}_{i=1}^{d_{A'}}\beta_{i}F_{i}$
where $\alpha_{i},\,\beta_{i}\geq0$.

{\small
\begin{eqnarray*}
\mbox{Tr}(W_{A'ABB'}\sigma_{A'ABB'}) & = & \mbox{Tr}(\mathrm{P}_{A'}\otimes\mathcal{W}_{AB}\otimes\mathrm{P}_{B'}\sigma_{A'ABB'})\\
 & = & \mbox{Tr}(\sum_{i=1}^{d_{A'}}\lambda_{i}E_{i}\otimes\mathcal{W}_{AB}\otimes\sum_{j=1}^{d_{B'}}\beta_{j}F_{j}\sigma_{A'ABB'})\\
 & = & \sum_{i=1}^{d_{A'}}\sum_{j=1}^{d_{B'}}\alpha_{i}\beta_{j}\mbox{Tr}(E_{i}\otimes\mathcal{W}_{AB}\otimes F_{i}\sigma_{A'ABB'})\\
 & = & \sum_{i=1}^{d_{A'}}\sum_{j=1}^{d_{B'}}\alpha_{i}\beta_{j}\mbox{Tr}(\left(\mathrm{\mathbb{I}}_{A'}\otimes\mathcal{W}_{AB}\otimes\mathbb{I}_{B'}\right)\left(E_{i}\otimes\mathrm{\mathbb{I}}_{AB}\otimes F_{j}\right)\sigma_{A'ABB'}\left(E_{i}\otimes\mathrm{\mathbb{I}}_{AB}\otimes F_{j}\right))\\
 & = & \sum_{i=1}^{d_{A'}}\sum_{j=1}^{d_{B'}}\alpha_{i}\beta_{j}\mbox{Tr}(\mathcal{W}_{AB}\text{Tr}_{A'B'}\left[\left(E_{i}\otimes\mathrm{\mathbb{I}}_{AB}\otimes F_{j}\right)\sigma_{A'ABB'}\left(E_{i}\otimes\mathrm{\mathbb{I}}_{AB}\otimes F_{j}\right)\right])\\
 & = & \sum_{i=1}^{d_{A'}}\sum_{j=1}^{d_{B'}}\alpha_{i}\beta_{j}\mbox{Tr}(\mathcal{W}_{AB}\sigma_{AB}^{i,j})\geq0,
\end{eqnarray*}
}where $\sigma_{AB}^{i,j}$ are some {\small separable} states in $\mathcal{S}(\mathbb{C}^{d_{A}}\otimes\mathbb{C}^{d_{B}})$.  In the next to the last line we use the fact that local operations do not change property of separability.
Which ends the proof.

Recall that  entanglement witnesses $\mathcal{W}_{AB}$ are decomposable
if $\mathcal{W}_{AB}=Q_{1}+Q_{2}^{\Gamma}$, where $Q_{1},\, Q_{2}\geq0$
and $Q^{\Gamma}$ denotes the partail transposition, otherwis we say that they are  indecomposable. It is clear from
the definition that a decomposable EW cannot detect an entangled $\text{PPT}$ 
state, therefore such EW is useless in the search for bound entangled states.
\begin{theorem} If $\mathcal{W}_{AB}$ is indecomposable entanglement
witness, then so is $\mathcal{W}_{A'ABB'}$ . \end{theorem}

$\mathit{Proof.}$ If $\mathcal{W}_{AB}$ is indecomposable entanglement
witness then there exists the state $\rho_{AB}\in\mathcal{S}(\mathbb{C}^{d_{A}}\otimes\mathbb{C}^{d_{B}})$
PPT such that $\text{{Tr}}\left(\mathcal{W}_{AB}\rho_{AB}\right)<0$.
Let us consider the extended state $\widetilde{\rho}_{A'ABB'}=\widetilde{\mathrm P}_{A'}\otimes\rho_{AB}\otimes\widetilde{\mathrm P}_{B'}\mathcal{\in S}(\mathbb{C}^{d_{A'}d_{A}}\otimes\mathbb{C}^{d_{B}d_{B'}})$,
where $\widetilde{\mathrm{P}}_{A'}\in\mathcal{B}(\mathbb{C}^{d_{A'}})$ and
$\widetilde{\mathrm{P}}_{B'}\in\mathcal{B}(\mathbb{C}^{d_{B'}})$ are positive
semidefinite operators. To proof this theorem  we use the below fact

\begin{fact} If $\left(\mathbb{I}_{d_{A}}\otimes{\rm T}_{d_{B}}\right)\rho_{AB}\geq0$
, then $\left(\mathbb{I}_{d_{A'}d_{A}}\otimes{\rm T}_{d_{B}d_{B'}}\right)\widetilde{\rho}_{A'ABB'}\geq0$.\end{fact} To prove this fact we use decomposition of $\rho_{AB}=\sum_{i,j=1}^{d_{A}}e_{ij}\otimes\rho_{ij}$.
If $\rho_{AB}^{\Gamma}=\left(\mathbb{I}_{d_{A}}\otimes{\rm T}_{d_{B}}\right)\rho_{AB}=\sum_{i,j=1}^{d_{A}}e_{ij}\otimes\rho_{ij}^{T}\geq0$
then we have 

\begin{eqnarray*}
\rho_{A'ABB'}^{\Gamma} & = & \left(\mathbb{I}_{d_{A'}d_{A}}\otimes{\rm T}_{d_{B}d_{B'}}\right)\widetilde{\rho}_{A'ABB'}\\
 & = & \left(\mathbb{I}_{d_{A'}d_{A}}\otimes{\rm T}_{d_{B}d_{B'}}\right)\widetilde{\mathrm P}_{A'}\otimes\sum_{i,j=1}^{d_{A}}e_{ij}\otimes\rho_{ij}\otimes\widetilde{\mathrm P}_{B'}\\
 & = & \widetilde{\mathrm P}_{A'}\otimes\sum_{i,j=1}^{d_{A}}e_{ij}\otimes\rho_{ij}^{T}\otimes\widetilde{\mathrm P}_{B'}^{T}\\
 & = & \widetilde{\mathrm P}_{A'}\otimes\rho_{AB}^{\Gamma}\otimes\widetilde{\mathrm P}_{B'}^{T}\geq0,
\end{eqnarray*}
because $\widetilde{\mathrm P}_{B'}^{T}\geq0,$ and we obtain

\[
\text{Tr}\left(\mathcal{W}_{AABB'}\rho_{A'ABB'}\right)=\text{Tr}\left(\mathrm{P}_{A'}\widetilde{\mathrm{P}}_{A'}\right)\text{Tr}\left(\mathcal{W}_{AB}\rho_{AB}\right)\text{Tr}\left(\mathrm{P}_{B'}\widetilde{\mathrm{P}}_{B'}\right)<0,
\]
which means that $\mathcal{W}_{AABB'}$  is an indecomposable entanglement witness. Which ends the proof.

Given an entanglement witness $\mathcal{W}_{AB}$
one defines a set of all entangled states in $\mathcal{H}_{AB}$ detected
by $\mathcal{W}_{AB}$
\[
D_{\mathcal{W}_{AB}}=\left\{ \rho_{AB}\in\mathcal{S}\left(\mathcal{H}_{AB}\right)|\text{Tr}\left(\rho_{AB}\mathcal{W}_{AB}\right)<0\right\} .
\]
Suppose now that we are given two entanglement witnesses $\mathcal{W}_{1}$
and $\mathcal{W}_{2}$ in $\mathcal{H}_{AB}$.
\begin{definition}We cal $\mathcal{W}_{1}$ finer than $\mathcal{W}_{2}$
if $D_{\mathcal{W}_{1}}\supseteq D_{\mathcal{W}_{2}}$. $\mathcal{W}$
is called optimal if there is no other entanglement witness which
is finer than $\mathcal{W}$.\end{definition} It is clear that optimal
witnesses are sufficient to detect all entangled states.
The Authors of Ref. \citep{O} formulated the following sufficient
condition for the optimality. For a given entanglement witness of
$\mathcal{W}_{AB}\in\mathcal{B}(\mathbb{C}^{d_{A}}\otimes\mathbb{C}^{d_{B}})$ we define the set 
\[
P_{\mathcal{W}_{AB}}=\left\{ \left|\phi_{A}\otimes\psi_{B}\right\rangle \in\mathcal{H}_{A}\otimes\mathcal{H}_{B}|\text{Tr}\left(\mathcal{W}_{AB}\left|\phi_{A}\otimes\psi_{B}\left\rangle \right\langle \phi_{A}\otimes\psi_{B}\right|\right)=0\right\} .
\]
We say that $\mathcal{W}_{AB}$ possesses a spanning property if $\text{span}P_{\mathcal{W}}=\mathcal{H}_{AB}.$

\begin{proposition}\label{prop:1}\citep{O}Any entanglement witness possessing
a spanning property is optimal. \end{proposition} Note that the spanning
property is not a necessary condition for optimality of the EW since
the Choi map \citep{Choi 1,choi-2}, gives rise to an optimal entanglement
witness that has no spanning property.

Consider now an indecomposable EW $\mathcal{W}_{AB}$
  and define a set
\[
D_{\mathcal{W}_{AB}}^{PPT}=\left\{ \rho_{AB}\in\mathcal{S}\left(\mathcal{H}_{AB}\right)|\text{Tr}\left(\rho_{AB}\mathcal{W}_{AB}\right)<0\ \text{and}\ \ensuremath{\rho_{AB}^{\Gamma}\geq0}\right\} ,
\]
 i.e. a set of PPT entangled states detected by $\mathcal{W}_{AB}$
 .

\begin{definition}We call $\mathcal{W}_{1}$
  nd-finer than $\mathcal{W}_{2}$
  if  $D_{\mathcal{W}_{1}}^{PPT}\supseteq D_{\mathcal{W}_{2}}^{PPT}
 .$  $ \mathcal{W}$
  is called nd-optimal if there is no other entanglement witness which is nd-finer than $\mathcal{W}$
 .\end{definition} 
We see that nd-optymality is stronger than optimality.

\begin{proposition}\label{prop:2}\citep{O} An entanglement witness is nd-optimal if and only if both $\mathcal{W}_{AB}$
  and $\mathcal{W}_{AB}^{\Gamma}$
  are optimal. \end{proposition}

Endowed with this information we can formulate the following theorem.

\begin{theorem} If $\mathcal{W}_{AB}\in\mathcal{B}\left(\mathcal{H}_{A}\otimes\mathcal{H}_{B}\right)$
(respectively $\mathcal{W}_{AB}^{\Gamma}\in\mathcal{B}\left(\mathcal{H}_{A}\otimes\mathcal{H}_{B}\right)$)
entanglement witness possess a spanning property, then so is $\mathcal{W}_{A'ABB'}\mathcal{\in B}\left(\mathcal{H}_{A'A}\otimes\mathcal{H}_{BB'}\right)$
(respectively $\mathcal{W}_{A'ABB'}^{\Gamma}\in\mathcal{B}\left(\mathcal{H}_{A'A}\otimes\mathcal{H}_{BB'}\right)$).
\end{theorem}
$\mathit{Proof.}$We proof this theorem in two steps:
\begin{enumerate}
\item {If $\mathcal{W}_{AB}$  possess a spanning property, then so is $\mathcal{W}_{A'ABB'} \, (see\, \eqref{eq:witness}).$

We know that there exists the set of vectors \[\left\{ \left|\phi_{A}\otimes\psi_{B}\right\rangle \in\mathcal{H}_{A}\otimes\mathcal{H}_{B}|\text{Tr}\left(\mathcal{W}_{AB}\left|\phi_{A}\otimes\psi_{B}\left\rangle \right\langle \phi_{A}\otimes\psi_{B}\right|\right)=0\right\} \]
spans Hilbert space $\mathcal{H}_{A}\otimes\mathcal{H}_{B}$, because
$\mathcal{W}_{AB}$  possess a spanning property. Let $\left\{ \left|e_{A'}\right\rangle \right\} $
 denote an orthonormal basis in $\mathcal{H}_{A'}$ and respectively
$\left\{ \left|f_{B'}\right\rangle \right\} $ in $\mathcal{H}_{B'}$,
then we can easily construct the new set $\left\{ \left|e_{A'}\otimes\phi_{A}\otimes\psi_{B}\otimes f_{B'}\right\rangle \right\} $
which spans Hilbert space $\mathcal{H}_{A'A}\otimes\mathcal{H}_{BB'}$.
Now we need to show that for any vector from this set and $\mathcal{W}_{A'ABB'}$
the condition $\text{Tr}\left(\mathcal{W}_{A'ABB'}\left|e_{A'}\otimes\phi_{A}\otimes\psi_{B}\otimes f_{B'}\left\rangle \right\langle e_{A'}\otimes\phi_{A}\otimes\psi_{B}\otimes f_{B'}\right|\right)=0$ is satisfied. Let us denote} $E=\left|e_{A'}\left\rangle \right\langle e_{A'}\right|$,
$F=\left|f_{B'}\left\rangle \right\langle f_{B'}\right|$ and $\mathrm{P}_{\phi\otimes\psi}=\left|\phi_{A}\otimes\psi_{B}\left\rangle \right\langle \phi_{A}\otimes\psi_{B}\right|${\small .
Indeed, we have
\[
\text{Tr}\left(\mathcal{W}_{A'ABB'}E\otimes\mathrm{P}_{\phi\otimes\psi}\otimes F\right)=\text{Tr}\left(\mathrm{P}_{A'}E\right)\text{Tr}\left(\mathcal{W}_{AB}\mathrm{P}_{\phi\otimes\psi}\right)\text{Tr}\left(\mathrm{P}_{B'}F\right)=0,
\]
}it holds because of property of  $\mathcal{W}_{AB}$.

\item {If $\mathcal{W}_{AB}^{\Gamma}$  possess a spanning property, then so is $\mathcal{W}_{A'ABB'}^{\Gamma}.$ To prove this we use the same way like before. We know
that there exist the set of vectors 
\[
\left\{ \left|\tilde{\phi}_{A}\otimes\tilde{\psi}_{B}\right\rangle \in\mathcal{H}_{A}\otimes\mathcal{H}_{B}|\text{Tr}\left(\mathcal{W}_{AB}^{\Gamma}\left|\tilde{\phi}_{A}\otimes\tilde{\psi}_{B}\left\rangle \right\langle \tilde{\phi}_{A}\otimes\tilde{\psi}_{B}\right|\right)=0\right\} 
\]
 spans Hilbert space $\mathcal{H}_{A}\otimes\mathcal{H}_{B}$, because
$\mathcal{W}_{AB}^{\Gamma}$  possess a spanning property. Let $\left\{ \left|\tilde{e}_{A'}\right\rangle \right\} $
 denote an orthonormal basis in $\mathcal{H}_{A'}$ and respectively
$\left\{ \left|\tilde{f}_{B'}\right\rangle \right\} $ in $\mathcal{H}_{B'}$,
then we can construct the new set $\left\{ \left|\tilde{e}_{A'}\otimes\tilde{\phi}_{A}\otimes\tilde{\psi}_{B}\otimes\tilde{f}_{B'}\right\rangle \right\} $
which spans Hilbert space $\mathcal{H}_{A'A}\otimes\mathcal{H}_{BB'}$.
We need to show that for any vector from this set and $\mathcal{W}_{A'ABB'}^{\Gamma}$
the condition $\text{Tr}\left(\mathcal{W}_{A'ABB'}^{\Gamma}\tilde{E}\otimes\mathrm{\tilde{P}}_{\phi\otimes\psi}\otimes\mathrm{\tilde{F}}\right)=0$
is satisfied{\small , where} $\tilde{E}=\left|\tilde{e}_{A'}\left\rangle \right\langle \tilde{e}_{A'}\right|$,
$\tilde{F}=\left|\tilde{f}_{B'}\left\rangle \right\langle \tilde{f}_{B'}\right|$
and $\mathrm{\tilde{P}}_{\phi\otimes\psi}=\left|\tilde{\phi}_{A}\otimes\tilde{\psi}_{B}\left\rangle \right\langle \tilde{\phi}_{A}\otimes\tilde{\psi}_{B}\right|.$
{\small 
\begin{eqnarray*}
\text{Tr}\left(\mathcal{W}_{A'ABB'}^{\Gamma}\tilde{E}\otimes\mathrm{\tilde{P}}_{\phi\otimes\psi}\otimes\mathrm{\tilde{F}}\right) & = & \text{Tr}\left(\left(\mathrm{P}_{A'}\otimes\mathcal{W}_{AB}^{\Gamma}\otimes\mathrm{P}_{B'}^{T}\right)\tilde{E}\otimes\mathrm{\tilde{P}}_{\phi\otimes\psi}\otimes\mathrm{\tilde{F}}\right)\\
 & = & \text{Tr}\left(\mathrm{P}_{A'}\tilde{E}\right)\text{Tr}\left(\mathcal{W}_{AB}^{\Gamma}\mathrm{\tilde{P}}_{\phi\otimes\psi}\right)\text{Tr}\left(\mathrm{P}_{B'}^{T}\tilde{F}\right)=0
\end{eqnarray*}
}we use the fact that $\mathcal{W}_{AB}^{\Gamma}$  possesses a spanning property.}
\end{enumerate}
Theorem above says that if $\mathcal{W}_{AB}$ (respectively $\mathcal{W}_{AB}^{\Gamma}$) possesses spanning property then $\mathcal{W}_{A'ABB'}$ (respectively $\mathcal{W}_{A'ABB'}^{\Gamma}$) possesses this property and hence is optimal (nd-optimal) due to proposition \ref{prop:1} and proposition \ref{prop:2}.

\begin{theorem}(nontrivial extension) There are the states $\rho_{A'ABB'}\in S(\mathbb{C}^{d_{A'}d_{A}}\otimes\mathbb{C}^{d_{B'}d_{B}})$
which are detected by $\mathcal{W}_{A'ABB'}$, but the states $S(\mathbb{C}^{d_{A}}\otimes\mathbb{C}^{d_{B}})\ni\rho_{AB}=\text{{Tr}}_{A'B'}\left(\rho_{A'ABB'}\right)$
are not detected by $\mathcal{W}_{AB}$. \end{theorem}

$\mathit{Proof.}$ We will give explicit example of these states. Let
us consider Choi witness of the form:

\begin{equation}
\mathcal{W}_{AB}=\left(\begin{array}{ccc|ccc|ccc}
1 & 0 & 0 & 0 & -1 & 0 & 0 & 0 & -1\\
0 & 0 & 0 & 0 & 0 & 0 & 0 & 0 & 0\\
0 & 0 & 1 & 0 & 0 & 0 & 0 & 0 & 0\\
\hline 0 & 0 & 0 & 1 & 0 & 0 & 0 & 0 & 0\\
-1 & 0 & 0 & 0 & 1 & 0 & 0 & 0 & -1\\
0 & 0 & 0 & 0 & 0 & 0 & 0 & 0 & 0\\
\hline 0 & 0 & 0 & 0 & 0 & 0 & 0 & 0 & 0\\
0 & 0 & 0 & 0 & 0 & 0 & 0 & 1 & 0\\
-1 & 0 & 0 & 0 & -1 & 0 & 0 & 0 & 1
\end{array}\right).
\end{equation}
We extend this witness to the form (for convenience we will take $d_{A'}=1,$
$d_{B'}=2$) $\mathcal{W}_{ABB'}=\mathcal{W}_{AB}\otimes \mathrm{P}_{B'}$,
where $\mathrm{P}_{B'}\in\mathcal{B}(\mathbb{C}^{2})$. Let us consider the state $\rho_{ABB'}\in\mathcal{S}(\mathbb{C}^{3}\otimes\mathbb{C}^{3}\otimes\mathbb{C}^{2})\simeq\mathcal{S}(\mathbb{C}^{3}\otimes\mathbb{C}^{6})$
of the following form:

\begin{equation}
\rho_{ABB'}=\frac{1}{(a_{11}+a_{22}+b_{11}+b_{22})}{\displaystyle \sum_{i,j=1}^{3}\left|i\left\rangle \right\langle j\right|\otimes\rho_{ij}},
\end{equation}
where:

\begin{equation}
\rho_{ii}=\left(S^{i-1}\otimes\mathbb{I}_{2}\right)X\left(S^{i-1}\otimes\mathbb{I}_{2}\right)^{\dagger},
\end{equation}
$S$ is the shift operator defined by $S\left|k\right\rangle =\left|k+1\right\rangle ,$
and 
\begin{equation}
X=\left(\begin{array}{cc|cc|cc}
a_{11} & a_{12} & 0 & 0 & 0 & 0\\
a_{21} & a_{22} & 0 & 0 & 0 & 0\\
\hline 0 & 0 & 0 & 0 & 0 & 0\\
0 & 0 & 0 & 0 & 0 & 0\\
\hline 0 & 0 & 0 & 0 & b_{11} & b_{12}\\
0 & 0 & 0 & 0 & b_{21} & b_{22}
\end{array}\right),
\end{equation}
and

\begin{equation}
\rho_{ij}=\left|i\left\rangle \right\langle j\right|\otimes\left(\begin{array}{cc}
a_{11} & a_{12} \\
a_{21}  & a_{22} 
\end{array}\right),\qquad i\neq j.
\end{equation}
The state $\rho_{ABB'}$ is positive semidefinite if and only if $a=\left(\begin{array}{cc}
a_{11} & a_{12}\\
a_{21} & a_{22}
\end{array}\right)\geq0$ and $b=\left(\begin{array}{cc}
b_{11} & b_{12}\\
b_{21} & b_{22}
\end{array}\right)\geq0$. It is easy to show that

\begin{equation}
\text{{Tr}}\left(\mathcal{W}_{ABB'}\rho_{ABB'}\right)=\frac{3}{\left(\text{{Tr}}\left(a+b\right)\right)}\text{{Tr}}\left(\mathrm{P}_{B'}\left(b-a\right)\right)
\end{equation}
and
\begin{equation}
\text{{Tr}}\left(\mathcal{W}_{AB}\rho_{AB}\right)=\frac{3}{\left(\text{{Tr}}\left(a+b\right)\right)}\text{{Tr}}\left(b-a\right),
\end{equation}
where $\rho_{AB}=\text{{Tr}}_{B'}(\rho_{ABB'})\in\mathcal{S}(\mathbb{C}^{3}\otimes\mathbb{C}^{3})$.
Let us choose for example $\mathrm{P}_{B'}=\left(\begin{array}{cc}
1 & 1\\
1 & 1
\end{array}\right)$, $a_{21}=a_{12}$ $\left(b_{21}=b_{12}\right)$ and $b_{11}=a_{11}$
and $b_{22}=a_{22}$. For this particular state we get $\text{{Tr}}\left(\mathcal{W}_{ABB'}\rho_{ABB'}\right)<0$
if $b_{12}\neq a_{12}$ and $\text{{Tr}}\left(\mathcal{W}_{AB}\rho_{AB}\right)=0$ (In fact the state $\rho_{AB}$ is separable if and only if $a_{11}=0$ and $a_{22}=0$).
Which ends the proof.

\section{Extended entanglement witnesses and measurement  device independent entanglement witnesses}

Here we shall show that there is a natural connection of an extended EW of the form $\mathcal{W}_{A'ABB'}=\mathrm{P}_{A'}\otimes\mathcal{W}_{AB}\otimes\mathrm{P}_{B'},$
for $\mathrm{P}_{A'},\:\mathrm{P}_{B'}\geq0$ and the measurement
device independent entanglement witnesses (MDIEW). First we shall recall the
scheme for measurement device independent entanglement witnesses in the spirit
of \citep{Gisin} which is built upon the previous idea of Buscemi
\citep{Buscemi}. 

 Let us recall the concept of MDIEW represents an operator, or in broader
sense quantum operation that allows us to detect entanglement of a
given quantum state $\rho_{AB}$ in the case when (i) the measurement device is 
not fully controlled and may be misaligned
(ii) even the dimensionality 
of Hilbert space $\mathcal{H}_{AB}=\mathcal{H}_{A}\otimes\mathcal{H}_{B}$
associated with the state $\rho_{AB}$ is uncontroled. 
The point (ii) should be understood in the sense that 
the experimentalists assume that there is some Hilbert space 
associated with the state $\rho_{AB}$; however, due to misalignment 
of the measurement device it hapens that it may couple 
with other degrees of freedom of the states.
The extremal, purely hypotetical example would be the situation when 
Alice and Bob expect to measure polarisation entanglement of the 
two photons, but their local measurement devices in reality are 
coupled to other degree of freedom like angular momentum or 
frequency, and neither of the observers knows about it. 
In this situation, there is  danger that the physical state of 
the new (angular momentum) degrees of freedom is separable, but due to the nature 
of the coupling eventually the false information about the presence of entanglement might be reported
in the sense that the overall statistical mean value might be negative. 
The MDIEW-s are just designed to avoid this situation. They must not report 
false entanglement if the actual (not hypothetical, or expected one)
physical state is separable. 
The needed element of the corresponding action of the observers is a full
control of some degrees of freedom of local apparatus or systems.
In Figure 1, we have the setup in which local by controlled objects are
the prepared states $\sigma_{A'}^{s}\:$$\sigma_{B'}^{t}$ their Hilbert
spaces are known to Alice and Bob also in the sense that local POVM's
(represented in Figure 1 as boxes) couple the local parts of the system
in the state $\rho_{AB}$ exactly to the degrees of freedom corresponding
to the known Hilbert spaces $\mathcal{H}_{A'}$ and $\mathcal{H}_{B'}$.
However, the nature of this coupling is asymetric: 
we know that it couples something known (physical degrees associated 
with the known states $\sigma_{A'}^{s}\:$$\sigma_{B'}^{t}$) to those
(associated with the physical degrees of the state $\rho_{AB}$)  
that are assumed by the observers to be the same as the previous ones
but in practice it does not need to be so.
In that sense the observers expect that 
$\mathcal{H}_{A'}$ and $\mathcal{H}_{B'} \cong \mathcal{H}_{A}$ and $\mathcal{H}_{B}$
in the physical sense, ie. they describe systems with the same 
degrees of freedom eg. photon polarisation while this may not be true - 
this corresponds to the property (ii) above. 
The property (i) which is direct misalignment, says that 
it may even happen that the assumption abaut the physical character 
of the degrees of freedom of the state $\rho_{AB}$ may be correct
(i.e. they may really be polarized, as the observers expect);
however, the coupling with the polarization of the states $\sigma_{A'}^{s}\:$$\sigma_{B'}^{t}$
may be not good. For instance this may happen when the Hong-Ou-Mandel (H-O-M) interferometer 
coupling the polarisation of local Alice photon $A'$ with the polarisation of the incoming photon $A$ works badly.
For completness, let us recall here that double click in the standard H-O-M interferometer 
results in the event, when the two incoming polarizations of the two photons were 
projected jointly onto the maximally entangled state $\Psi_{-}=\frac{1}{\sqrt{2}} ( |01\rangle - |10\rangle)$
which, however, with the help of the polarization rotators may be modified 
to encompass projection onto the entangled state $\Psi_{+}=\frac{1}{\sqrt{2}} ( |00\rangle + |11\rangle)$
(see (\cite{Ekert} for the experimental application of the interferometer in the experiment 
inplementing nonlinear entanglement witness associated with entropic inequalities).

Summarising, the observers must control the preparation of  the 
states $\sigma_{A'}^{s}\:$$\sigma_{B'}^{t}$, and be sure that 
that comes into the local device and is labeled by $A$ and $B$ is 
coupled just to those states.

This is the control that allows us to detect entanglement of $\rho_{AB}$
in the measurement device independent i.e. in such a way that 
false report about entanglement will never occur.

\begin{figure}
\includegraphics[scale=0.45]{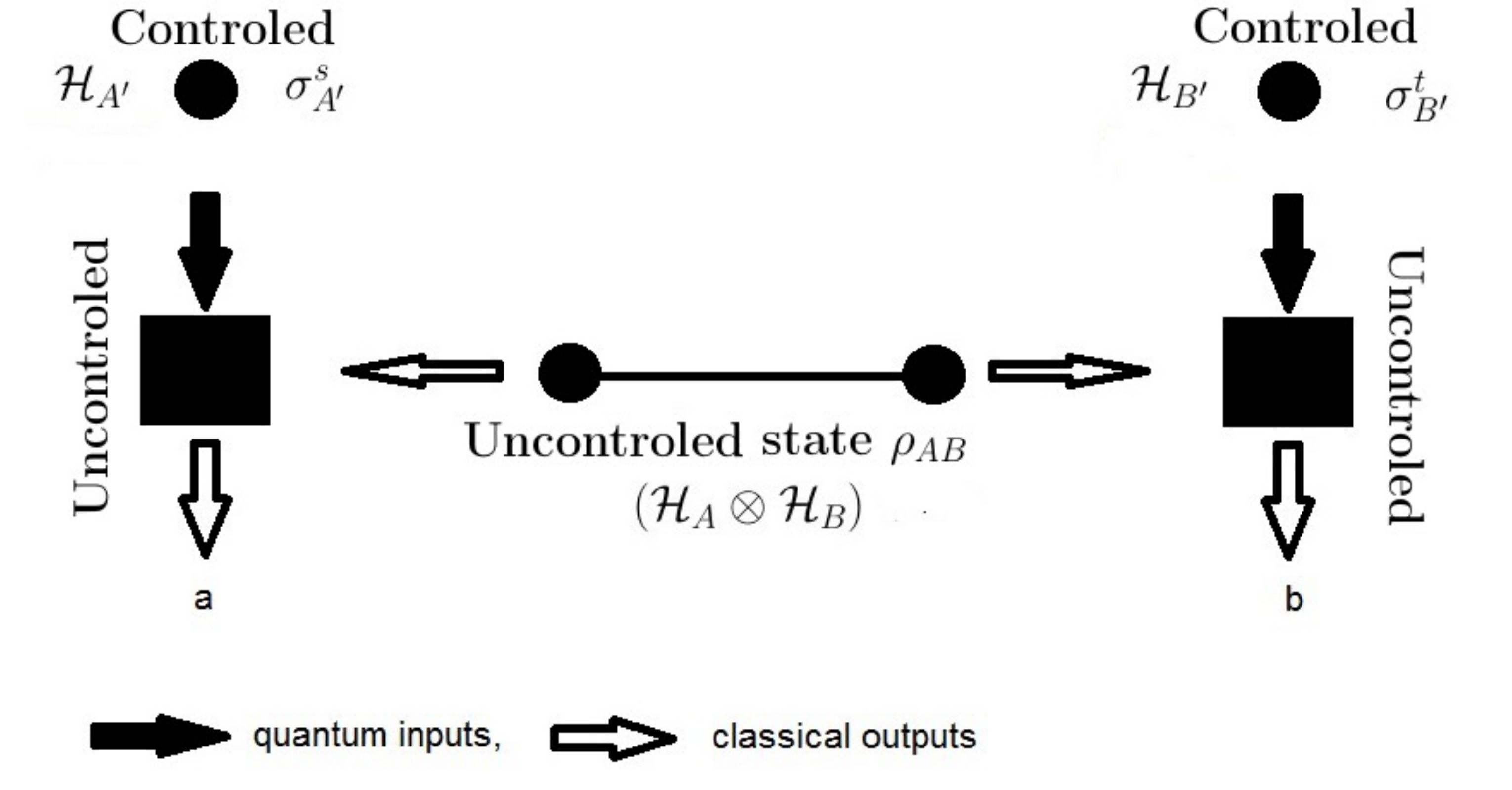}\caption{
The idea of the measurement device independent entanglement witness (MDIEW), is proposed in \citep{Gisin}. The observers control 
the local input states; however, they may not control how they couple to the 
incoming particles in state $\rho_{AB}$. Even the physical degrees of freedom of the latter 
may be completely different than expected by the observers. 
Nevertheless, the false report of the entanglement of the state $\rho$ {\it never} takes place 
which represens the essence of the  MDIEW concept.}
\end{figure}

The above description was presented  also in a more mathematical way \citep{Gisin}.
To recall the description suppose that there is a witness $W_{AB}$ which detects entanglement
of the state $\rho_{AB}$. The crucial idea of \citep{Gisin} is that
to represent the witness as a linear combination of quantum states
from Figure 1, namely:

\begin{equation}
W_{AB}={\displaystyle \sum_{s,t}\beta_{st}\sigma_{A'}^{s\: T}\otimes\sigma_{B'}^{t\: T}}.\label{eq:Wstates}
\end{equation}
Suppose now, that the observers use the measuring devices with two
inputs $\mathcal{H}_{A}$, $\mathcal{H}_{A'}$ ( and $\mathcal{H}_{B}$,
$\mathcal{H}_{B'}$ respectively), and claim that the device performs
entangling the von Neuman measurements $\left\{ P_{A'A}^{+},\: P_{A'A}^{+\bot}\right\} $
and $\left\{ P_{BB'}^{+},\: P_{BB'}^{+\bot}\right\} $, where $P_{A'A}^{+}=\left|\Psi_{A'A}^{+}\left\rangle \right\langle \Psi_{A'A}^{+}\right|$
with $\left|\Psi_{A'A}^{+}\right\rangle ={\displaystyle \frac{1}{\sqrt{d_{A}}}\sum_{k=0}^{d_{A'}-1}\left|k\right\rangle \left|k\right\rangle }$,
and $P_{BB'}^{+}$ in full analogy. Note that if both observers report
the result $"0"$ which corresponds to local projections onto $P_{A'A}^{0} \equiv P_{A'A}^{+}$,
$P_{BB'}^{0}\equiv P_{BB'}^{+}$ then the joint probability with the inputs $\sigma^{s}$,
$\sigma^{t}$ is 

\begin{equation}
P\left(a=0,b=0\mid st\right)\overset{def}{=}\text{Tr}\left(\rho_{AB}\otimes\sigma_{A'}^{s\: T}\otimes\sigma_{B'}^{t\: T}P_{A'A}^{0}\otimes P_{BB'}^{0}\right).\label{eq:praw1}
\end{equation}
Combining this with the formula (\ref{eq:Wstates}) gives the mean
value of the entanglement witness

\begin{equation}
\mathcal{W}=\left\langle W_{AB}\right\rangle ={\displaystyle \sum_{s,t}\beta_{st}P\left(a=0,b=0\mid st\right)}.\label{eq:wit2}
\end{equation}
with the standard mean value $\left\langle W_{AB}\right\rangle=Tr(\rho_{AB} W_{AB})$ of the
witness observable $W_{AB}$ is defined by the formula (\ref{eq:Wstates}).
The major claim of the \citep{Gisin} is that instead of the intended von Neumann
measurements the local observers perform, due to misalignment, arbitrary POVM's $\left\{ \tilde{P}_{A'A}^{0},\: \tilde{P}_{A'A}^{1}\right\} $,
$\left\{ \tilde{P}_{BB'}^{0},\: \tilde{P}_{BB'}^{1}\right\} $ ($0 \leq \tilde{P}_{A'A}^{0}\leq I_{A'A}$, 
$0 \leq \tilde{P}_{BB'}^{0}\leq I_{BB'}$, $\tilde{P}_{A'A}^{0}+ \tilde{P}_{A'A}^{1}= I_{A'A}$,
$\tilde{P}_{BB'}^{0}+ \tilde{P}_{BB'}^{1}= I_{BB'}$)  and at the same time,
the dimensions of $\mathcal{H}_{A}$, $\mathcal{H}_{B}$ and the degrees of freedom 
they represent (associated with the state $\rho_{AB}$) 
are \textit{uncontroled}, then still the negative
value of the quantity 
\begin{equation}
\tilde{\mathcal{W}} \equiv {\displaystyle \sum_{s,t}\beta_{st}\tilde{P}\left(a=0,b=0\mid st\right)},\label{eq:wit2-1}
\end{equation}
reports entanglement of $\rho_{AB}$ (or, equivalently, the value is always nonegative for separable state $\rho_{AB}$),
if only Alice and Bob can fully control preparation of $\sigma_{i}^{s}$ and $\sigma_{i}^{t}$ 
which means that they control the physical degrees of freedom associated with $\mathcal{H}_{A'}$, $\mathcal{H}{}_{B'}$.
Here the conditional probability corresponds to the '0' outcomes (represented by the 
operators $\tilde{P}_{A'A}^{0}$, $\tilde{P}_{BB'}^{0}$ introduced above)
of some POVM's (may be uncontroled by the experimentalists) which couples the correctly prepared states
$\sigma_{A'}^{s}$$,\sigma_{B'}^{t}$ to the experimentally investigated state $\rho_{AB}$. 
Also it should be stressed that we denoted, in full analogy to the formula (\ref{eq:praw1})
\begin{equation}
\tilde{P}\left(a=0,b=0\mid st\right)\overset{def}{=}\text{Tr}\left(\rho_{AB}\otimes\sigma_{A'}^{s\: T}\otimes\sigma_{B'}^{t\: T}\tilde{P}_{A'A}^{0}\otimes \tilde{P}_{BB'}^{0}\right)\label{eq:praw2}
\end{equation}
with the only difference that  - unlike in (\ref{eq:praw1}) -  here the operators $\tilde{P}_{A'A}^{0}$,
$\tilde{P}_{BB'}^{0}$  are no longer projectors but just elements of some local POVM-s. 

%\label{eq:praw1}

Here we will interpret the result of \citep{Gisin} in terms of our
extension of entanglement witnesses. Namely we will show that the
quantity (\ref{eq:wit2-1}) is always positive for any separable states
$\rho_{AB}$ because then it can be actually represented as mean value
of the convex combination of some extended witnesses on the unnormalized
product states. 

Consider any separable state $\rho_{AB}={\displaystyle \sum}_{i}p_{i}\rho_{A}^{i}\otimes\rho_{B}^{i}$
(uncontroled). As a result of the previous section we
have the family of extended entanglement witnesses 

\[
W_{A'ABB'}^{i}=\rho_{A'}^{i}\otimes\mathcal{W}_{AB}\otimes\rho_{B'}^{i},
\]
with $\rho_{A'}^{i},\:\rho_{B'}^{i}\geq0$. As such they have positive
mean values on any product of two positive operators

\[
\text{Tr}\left(W_{A'ABB'}^{i}X_{A'A}\otimes Y_{BB'}\right)\geq0,
\]
with $X_{A'A},\: Y_{BB'}\geq0$. In particular we may put in their places the POVM elements 
considered above ie. assume $X_{A'A}=\tilde{P}_{A'A}^{0}$, $Y_{BB'}=\tilde{P}_{BB'}^{0}$, getting it this way
\[
\text{Tr}\left(W_{A'ABB'}^{i}\tilde{P}_{A'A}^{0}\otimes \tilde{P}_{BB'}^{0}\right)\geq0.
\]
Taking further (with help of the probabilities $\{ p_i \}$
defining separable $\rho_{AB}$) the convex combination and remembering the 
notation (\ref{eq:praw2}) it is easy to check that

\[
\tilde{\mathcal{W}}={\displaystyle \sum_{s,t}\beta_{st}\tilde{P}\left(a=0,b=0\mid st\right)=\sum_{i}p_{i}\text{Tr}\left(W_{A'ABB'}^{i}\tilde{P}_{A'A}^{0}\otimes \tilde{P}_{BB'}^{0}\right)\geq0.}
\]
This means that if such a convex combination were negative, then the
state $\rho_{AB}$ must have been entangled. This fact is a crucial 
property of MDIEW: since quantum entanglement is a resource we would not like 
be informed about its presence when actually it is not there and that is 
what MDIEW guarantees us to avoid.

\section{Conclusions }

We have provided product extension of entanglement witnesses and shown
that they inherit the properties of indecomposability and spanning property (hence optimality)
from the original entanglement witnesses. We also have shown that
the structure of such extension is naturally present in the mathematics
of the scheme of measurement device independent entanglement witnessing of a given
entanglement state. 
It should be stressed that the
general construction of product extension of the entanglement
witness and its relation to measurement device independent
entanglement witnesses can be easily generalised
to the case of multipartite case.
It is interesting to analyse the optimality
concept for the mulitpatite version, since it has not
been developed yet  (see \cite{Gisin}).

There is a natural question whether the concept presented here has any relation
to the general scheme of quantum cryptography which requires quantum
states with composed local systems (see \citep{key-3} and references
therein).

\section*{Acknowledgments}

This work was  supported by a postdoc internship (A. Rutkowski)
decision number DEC\textendash{} 2012/04/S/ST2/00002, from the Polish
National Science Center and EC project QUASAR (P. Horodecki). We thank Ryszard Horodecki and  Gniewomir Sarbicki for valuable discussion.

\bibliographystyle{plain}

\begin{thebibliography}{10}
\bibitem{QIT} M. A. Nielsen and I. L. Chuang, \textit{Quantum computation
and quantum information}, Cambridge University Press, Cambridge, 2000.

\bibitem{Horodecki-review} R. Horodecki, P. Horodecki, M. Horodecki
and K. Horodecki, Rev. Mod. Phys. \textbf{81}, 865 (2009).

\bibitem{HHH} M. Horodecki, P. Horodecki, and R. Horodecki, Phys.
Lett. A \textbf{223}, 1 (1996).

\bibitem{key-3}K. Horodecki, M. Horodecki, P. Horodecki, and J. Oppenheim
Phys. Rev. Lett. $\boldsymbol{94}$, 160502 (2005)

\bibitem{Terhal1} B. Terhal, Phys. Lett. A \textbf{271}, 319 (2000);
Linear Algebr. Appl. \textbf{323}, 61 (2000).

\bibitem{Terhal2} B. M. Terhal, Theor. Comput. Sci. \textbf{287},
313 (2002).


%================EWs


\bibitem{O} M. Lewenstein, B. Kraus, J. I. Cirac, and P. Horodecki,
Phys. Rev. A \textbf{62}, 052310 (2000).

\bibitem{Lew} M. Lewenstein, B. Kraus, P. Horodecki, and J. I. Cirac,
Phys. Rev. A \textbf{63}, 044304 (2001).

\bibitem{Lew2} B. Kraus, M. Lewenstein, and J. I. Cirac, Phys. Rev.
A \textbf{65}, 042327 (2002).

\bibitem{Lew3} P. Hyllus, O. G\"{u}hne, D. Bru\ss, and M. Lewenstein Phys.
Rev. A \textbf{72}, 012321 (2005).

\bibitem{Bruss} D. Bru\ss, J. Math. Phys. \textbf{43}, 4237 (2002).

\bibitem{Toth} G. T\'{o}th i O. G\"{u}hne, Phys. Rev. Lett. \textbf{94},
060501 (2005).

\bibitem{Bertlmann} R. A. Bertlmann, H. Narnhofer and W. Thirring,
Phys. Rev. A \textbf{66}, 032319 (2002).

\bibitem{Breuer} H.-P. Breuer, Phys. Rev. Lett. \textbf{97}, 080501
 (2006).

\bibitem{OSID} D. Chru\'{s}ci\'{n}ski and A. Kossakowski, Open Systems
and Inf. Dynamics, \textbf{14}, 275-294 (2007).

\bibitem{How} D. Chru\'{s}ci\'{n}ski and A. Kossakowski, J. Phys.
A: Math. Theor. \textbf{41} (2008) 145301; J. Phys. A: Math. Theor.
\textbf{41}  215201 (2008).

\bibitem{CMP} D. Chru\'{s}ci\'{n}ski and A. Kossakowski, Comm. Math.
Phys. \textbf{290}, 1051 (2009).

\bibitem{Gniewko} D. Chru\'{s}ci\'{n}ski, A. Kossakowski and G. Sarbicki,
Phys. Rev. A \textbf{80}, 042314 (2009).

\bibitem{iran-09} M.A. Jafarizadeh and N. Behzadi, Eur. Phys. J.
D \textbf{55}, 729 (2009).

\bibitem{iran} D. Chruscinski i A. Rutkowski, Eur. Phys. J. D \textbf{62},
273-277, (2011).
\bibitem{rut2} D. Chruscinski i A. Rutkowski,Open Systems
and Inf. Dynamics, \textbf{17}, 347-359 (2010).

\bibitem{Gisin} C. Branciard, D. Rosset, Y-Ch. Liang, and N. Gisin,
Phys. Rev. Lett. $\boldsymbol{110}$, 060405 (2013).

\bibitem{Buscemi} F. Buscemi, Phys. Rev. Lett. $\boldsymbol{108}$,
200401 (2012).

\bibitem{Choi 1}M.-D. Choi, Linear Algebra Appl. $\boldsymbol{12}$,
95 (1975).

\bibitem{choi-2} M.-D. Choi and T.-T. Lam, Math. Ann. $\boldsymbol{231}$,
1 (1977).

\bibitem{Bancal} J-D. Bancal, N. Gisin, Y-Ch. Liang, and S. Pironio
Phys. Rev. Lett. $\boldsymbol{106}$, 250404 (2011)

\bibitem{Ekert}
F. A. Bovino, G. Castagnoli, A. Ekert, P. Horodecki, C. M.
Alves, and A. V. Sergienko, Phys. Rev. Lett. {\bf 95}, 240407  (2005).
\end{thebibliography}

\end{document}